# Thermo-Optic Properties of Silicon-Rich Silicon Nitride for On-chip Applications

**Hani Nejadriahi,**[1, *] **Alex Friedman**[1], **Rajat Sharma**[2], **Steve Pappert**[1], **Yeshaiahu Fainman**[1], **and Paul Yu**[1]

[1]*Department of Electrical & Computer Engineering, University of California, San Diego, 9500 Gilman Drive, La Jolla, CA, 92093, USA*
[2]*VEO, 10509 Vista Sorrento Parkway, San Diego, CA, 92121*
*\*hnejadri@eng.ucsd.edu*

**Abstract:** We demonstrate the thermo-optic properties of silicon-rich silicon nitride (SRN) films deposited using plasma-enhanced chemical vapor deposition (PECVD). Shifts in the spectral response of Mach-Zehnder Interferometers (MZIs) as a function of temperature were used to characterize the thermo-optic coefficients of silicon nitride films with varying silicon contents. A clear relation is demonstrated between the silicon content and the exhibited thermo-optic coefficient in silicon nitride films, with the highest achievable coefficient being as high as $(1.65\pm0.08) \times 10^{-4}$ K$^{-1}$. Furthermore, we realize an SRN Multi-Mode Interferometer (MMI) based thermo-optic switch with over 20 dB extinction ratio and total power consumption for two-port switching of 50 mW.

## 1. Introduction

Presently, increasing the silicon content in silicon nitride thin films is being investigated towards creating a material platform that combines the benefits of silicon and stoichiometric silicon nitride. It has been shown that silicon-rich silicon nitride (SRN) films can demonstrate a highly enhanced nonlinear coefficient (third-order nonlinear susceptibility $\chi^{(3)}$ up to 4-5 times higher than that of crystalline silicon) while retaining low two-photon absorption in the Near Infra-Red (NIR) and a large transparency window such as that offered by stoichiometric silicon nitride [1,2]. As such, it is a viable candidate for on-chip applications such a super-continuum generation, wave-mixing, signal processing and light detection and ranging (LIDAR) [3-4]. The enhancement in SRN's nonlinear coefficient has commonly been attributed in the literature to its higher silicon content and in particular the presence of embedded silicon nanocrystals [5]. There has also been some discussion in literature on the possible enhancement of thermo-optic properties of silicon nitride as a function of its silicon content [6]. The thermo-optic coefficient of stoichiometric silicon nitride is commonly exploited for realizing on-chip devices such as electro-optic phase shifters [7] for various applications. However, on account of its low thermo-optic coefficient ($2.45\times10^{-5}$ K$^{-1}$) [8] and low refractive index, these devices suffer from low device efficiency in terms of high electrical power consumption and large device footprint [3,7]. Silicon on the other hand possesses a higher thermo-optic coefficient ($1.86\times10^{-4}$ K$^{-1}$), but suffers from a much narrower transparency window (transparent beyond 1.1 $\mu m$) and lower optical power handling capability on account of loss mechanisms such as two photon absorption (TPA) and surface carrier absorption (SCA) [5].

In this manuscript, we present SRN as an alternative material for on-chip applications that utilizes the thermo-optic effect. We provide a systematic evaluation of enhancement of PECVD deposited silicon nitride's thermo-optic coefficient as a function of its silicon content. Using this method, we conclusively demonstrate that the Si/N ratio of silicon nitride films can be adjusted by tuning the ratio of the precursor gases (SiH$_4$/N$_2$) in the PECVD system achieving films with a wide range of linear indices (1.92- 3.1, measured at $\lambda$ = 1550 nm). We demonstrate that highly silicon-rich silicon nitride films (refractive index, n > 3) possess a much wider transparency window than that of silicon and can be used to demonstrate low loss optical waveguiding in the C-band. We then characterize the thermo-optic response of MZIs fabricated with silicon nitride films of different compositions and quantitatively demonstrate an enhancement in the thermo-optic coefficient as a function of silicon content in these films. Finally, to validate our findings we employ the high linear index (3.1) SRN films with highly enhanced thermo-optic coefficient ($1.65\times10^{-4}$ $K^{-1}$) to realize an MMI switch with over 20 dB of extinction ratio with a port-to-port switching electrical power of 50 mW.

## 2. Film Preparations and optical properties

In the current work, PECVD was employed to deposit silicon nitride films of varying composition on top of oxide-on-silicon substrates. Deposition of silicon nitride using PECVD offers an advantage in terms of lower deposition temperature (350 °C in this study), as compared to LPCVD which is carried out at high temperatures (in the range of 700-800 °C). The ratio of precursor gases, $SiH_4$, $N_2$, and $NH_3$ was then varied, as shown in Table 1, to realize three samples, $S_1$, $S_2$, and $S_3$. For the purposes of ellipsometry measurements we maintained a film thickness of < 100 nm for all the three films. The Energy Dispersive Spectrum (EDS) measurements were carried out to estimate the ratio of silicon to nitrogen in these films and confirmed $S_1$ and $S_3$ to be the least and the most silicon rich films, respectively.

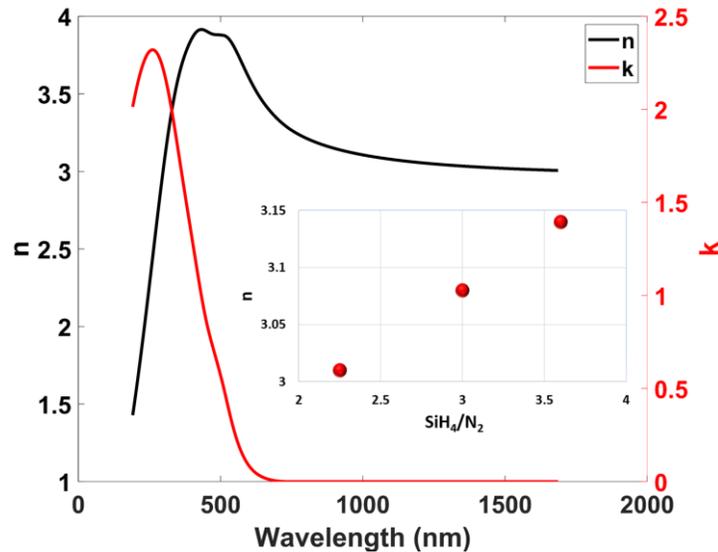

*Fig. 1. Ellipsometry measurements of the real and imaginary part of refractive index for SRN film of sample $S_3$ with n = 3.1 at λ=1550 nm; the inset shows the enhancement of the index of SRN films for the ammonia free condition with refractive indices of 3.01, 3.08, and 3.11.*

Ellipsometry measurements were then carried out to determine the real and imaginary parts of the linear refractive index of these films with the latter defining their transparency windows. Table 1 shows the real part of the refractive index measured at 1550 nm for films $S_1$ through $S_3$ confirming that linear refractive index increases with increase in silicon content in silicon nitride films. Fig.1 shows the dispersion in the real and imaginary part of the refractive index for sample $S_3$ over a wide wavelength range, 200-1600 nm. It is clearly visible that the imaginary part of the index, and hence the optical loss, of the film remains negligible even at wavelengths as low as 700 nm despite it having a silicon content significantly higher than that of stoichiometric silicon nitride. This also demonstrates that PECVD can be used to deposit highly silicon-rich silicon nitride films of index greater than 3 with low optical loss and wide transparency windows. Further, the inset in Fig.1 demonstrates how the refractive index of SRN films can be tuned as a function of the $SiH_4/N_2$ ratio in ammonia free recipes. Henceforth, for section 4 onwards, SRN refers to a film with refractive index of > 3.

**Table 1. Deposition parameters and the refractive index of silicon nitride with various compositions with thickness of 60 nm**

| Sample # | n ($\lambda$ =1550 nm) | SiH$_4$ Flow Rate (sccm) | N$_2$ Flow Rate (sccm) | NH$_3$ flow rate (sccm) |
|---|---|---|---|---|
| S$_1$ | 1.92 | 276 | 600 | 24 |
| S$_2$ | 2.25 | 400 | 200 | 10 |
| S$_3$ | 3.1 | 500 | 150 | 0 |

## 3. Loss characterization for SRN waveguides

While the ellipsometry measurements show that SRN films with linear refractive indices higher than 3 are indeed low loss in the NIR region (O and C bands), it is still important to evaluate their loss by in-waveguide propagation. For this purpose, we carried out the fabrication of bus-coupled ring resonators (i.e. all pass filter) with a range of coupling gaps using our PECVD deposited SRN films with n = 3.1 (measured at $\lambda$ = 1550 nm). Starting with a wafer with 3 $\mu m$ of thermal oxide on top of a silicon substrate, we first deposited a 320 nm thick SRN film using the recipe for the film sample S$_3$ (see Table 1). Bus-coupled ring resonators with waveguide widths of 400 nm and ring radius of 45 $\mu m$ are then fabricated using E-beam lithography followed by dry-etching as described in [9]. The device fabrication is then completed with a deposition of 2 $\mu m$ PECVD silicon dioxide as top clad followed by a 15-minute rapid thermal anneal (RTA) in a forming gas H$_2$/N$_2$ (10%:90%) ambient. The transmission response of these ring-resonators was then measured in the C-band using a tunable CW Agilent laser and a fiber-in, free-space out setup such as that presented in our previous work [9]. Fig. 2(a) and (b) show the measured response with a zoom-in of a single resonance and multiple resonances respectively for an SRN ring resonator with a 140 nm coupling gap for a propagating TM-polarized optical mode. The free-spectral range (FSR) measured using this response is 2.381nm, corresponding to a group index of 3.58, which in turn is very close to the value predicted using eigenmode analysis in Lumerical MODE with the same waveguide dimensions and measured refractive index. Fig. 2(a) also shows a single resonance from the response fit to a well-known Lorentzian expression [10] for the transmission response of a bus-coupled ring resonator which is then used to estimate the loss of the propagating mode. The optical loss coefficients extracted for the TE- and TM-like optical modes using this methodology are 7 dB/cm and 3.27 dB/cm respectively similar to those reported in [11]. These loss values are comparable to those obtained for deposited amorphous silicon waveguides [12] but higher than those usually observed for state-of-the-art silicon and silicon nitride waveguides [3,13]. However, it should be noted that this loss can be primarily attributed to an unoptimized fabrication process and not material absorption in the SRN film itself [14,15]. The combination of low optical loss and high refractive index would allow for the design of multi-mode straight and bent waveguides with comparable dimensions to those made using crystalline silicon [16].

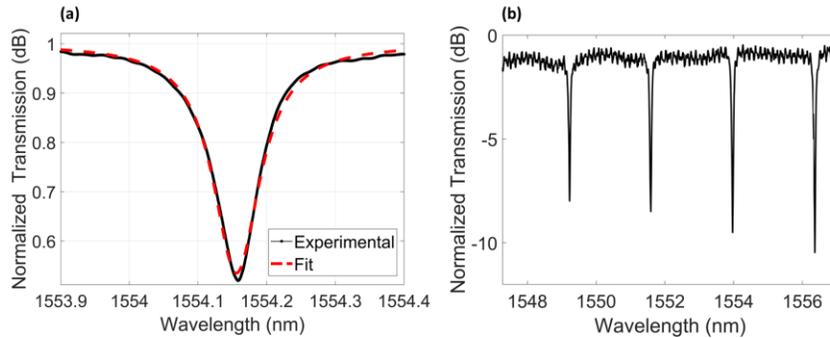

*Fig. 2. Passive transmission spectra for a ring resonator, corresponding to a 140 nm coupling gap and TM polarization for the optical mode (a) Single resonance at 1554.15 nm overlayed with the Lorenzian fit for the transmission response of a ring resonator (b) multiple resonances for the same ring resonator*

## 4. Thermo-optic characterization

In order to determine the effect of silicon content on the thermo-optic response of the silicon nitride films, we fabricated three different sets of MZIs ($S_1$, $S_2$, and $S_3$) using the three different films prepared following the deposition procedures described in section 2. Fig. 3(a) shows an optical microscope (OM) image of one of the fabricated MZI sets with integrated heater filaments and contact pads. Fig. 3(b) is a cross-sectional schematic of the device, showing the Ni:Cr heating filaments fabricated using a combination of photolithography (bilayer photoresist consisting of PMGI and AZ1512 and mask-less aligner Heidelberg 150) and RF sputtering (Denton 635). The dimensions of the heater filaments in each MZI set were 300 nm in height, 20 $\mu m$ in width, and four different lengths of 480, 980, 1480, and 1980 $\mu m$ as shown in Fig. 3 (a). On account of the high variation in the nitride film refractive indices, $n_{S1} = 1.92$, $n_{S2} = 2.25$, $n_{S3} = 3.1$, different film thicknesses and waveguide widths were chosen to ensure high overlap of the optical mode with the waveguide core. The waveguide dimensions for the three cases are listed in Table 2. Fig.3 (c) shows an SEM micrograph of a Y-splitter at the input end of one of the MZIs. It should be noted that while heater filaments were placed on both arms of the MZI, to balance the optical loss, one of the arms was left disconnected from the contact pads.

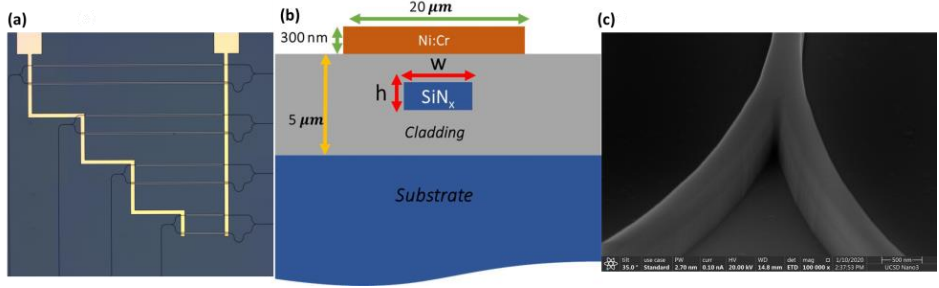

*Fig. 3. (a) OM top view of the four MZIs with integrated Ni:Cr heaters and Cr/Au contact pads consisting of 680 to 1920 µm of balanced lengths and 102 µm of imbalanced length  (b) schematic of the $SiN_x$ based MZIs with waveguide dimensions of h and w listed in Table 2 and total cladding thickness of 5 µm along with Ni:Cr heater filament of 20 µm wide and 300 nm tall (c) SEM image of the Y junction at the output port of the fabricated MZI*

To extract the thermo-optic coefficients of the three films, we then measured the spectral response of the shortest MZIs as a function of power in the heaters for all the three sets. Doing so allowed us to determine the change in the effective refractive index and hence the change in the material index of the waveguide core as a function of temperature. This step was performed by combining our experimental results with eigenmode and thermal modeling in Lumerical MODE and Lumerical DEVICE respectively. Transmission spectral scan were taken as a function of voltage applied to the heater. Fig. 4. (a) shows the experimentally measured spectral response of the shortest MZI in sample $S_3$ for a voltage range of 0 to 6 V. Fig. 4. (b) shows the extracted spectral shift $\Delta\lambda$, (in black) in the MZIs transmission response as a function of voltage applied. Also shown in the same figure (in red) is the experimentally measured power consumed in the heater. As shown, the spectral shift is a quadratic function of the voltage applied. This is because the power dissipated by the heater (P) is a quadratic function of the voltage applied ($P = \frac{V^2}{R}$), where R is the resistance of the heater element, and is linearly proportional to the change in temperature in the waveguide core ($\Delta T$). This was confirmed using thermal simulations carried out in Lumerical DEVICE. [8]. Eq. (1) describes the relationship between the experimentally observed $\Delta\lambda$ and the thermo-optically induced change in the effective index ($\Delta n_{eff}$) of the optical mode. Here $\lambda_{null}$ is the wavelength corresponding to the null point in transmission, $n_g$ is the group index calculated from the measured free spectral range (FSR) [10], $L_{mod}$ is the heater length of 480 $\mu m$, and $\Delta L = 102$ $\mu m$ is the imbalanced length.

$$\left(\frac{\Delta n_{eff}}{n_g}\right)\left(\frac{L_{mod}}{\Delta L}\right) = \frac{\Delta \lambda}{\lambda_{null}} \quad (1)$$

Hence knowing the spectral shift $\Delta\lambda$ as a function of change in temperature of the waveguide core ($\Delta T$), we can use Eq. (1) to calculate $\Delta n_{eff}$ as a function of $\Delta T$ shown in Fig. 4(c) for the three films. This change in effective index as a function of temperature, can then be related to the change in material indices via Eq. (2) where $\left(\frac{\partial n}{\partial T}\right)_{SiO_2} = 9 \times 10^{-6} K^{-1}$ [17,18] and $\left(\frac{\partial n}{\partial T}\right)_{SiNx}$ are the thermo-optic coefficients for the cladding oxide (as reported in literature) and silicon nitride respectively, and $\Gamma_{SiNx}$ and $\Gamma_{SiO_2}$ are the overlap factors of the mode with the silicon nitride waveguide core and the cladding oxide respectively. The overlap factor, $\Gamma$, in this equation is calculated in a manner similar to that reported in ref. [19,20].

$$\frac{\Delta n_{eff}}{\Delta T} = \Gamma_{SiO_2}\left(\frac{\partial n}{\partial T}\right)_{SiO_2} + \Gamma_{SiNx}\left(\frac{\partial n}{\partial T}\right)_{SiNx} \quad (2)$$

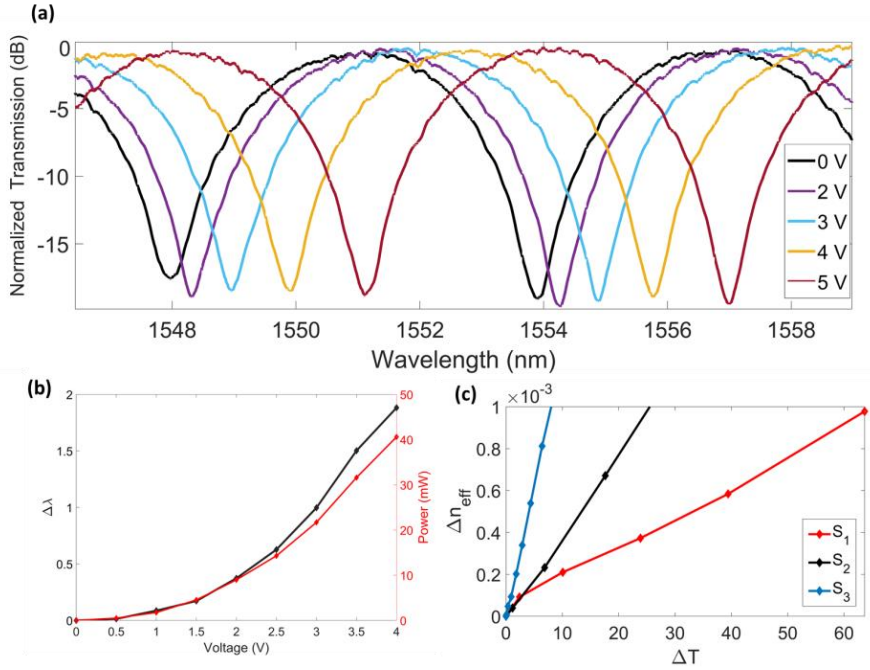

*Fig. 4. (a) Spectral shift of the null point in tansmission due to the applied voltage for sample $S_3$ (b) relative change in the wavelength and the power increase in the heater as functions of the applied voltage to the heater corresponding to (a), (c) the effective index change obtained from Eq. (1) for MZIs of all the three sets*

The calculated values of the overlap factors within the waveguide cores are shown in Table 2 below. Using these confinement factors, the $\frac{\Delta n}{\Delta T}$ for all three films was obtained. Thermo-optic coefficient extracted was determined to be $(1.65\pm0.08) \times 10^{-4}$ K$^{-1}$ for $S_3$, $(5.85\pm0.5) \times 10^{-5}$ K$^{-1}$ for $S_2$, and $(1.82\pm0.8) \times 10^{-5}$ K$^{-1}$ for $S_1$. The uncertainties in the coefficients arise from the shape and the thickness of the cladding oxide. The enhancement of the coefficient can be seen in the plot below as a function of the measured refractive index values for each of them. The result clearly shows the enhancement as a function of increase in the silicon content of the silicon nitride films with the highest achieved thermo-optic coefficient being comparable to that of crystalline silicon [21].

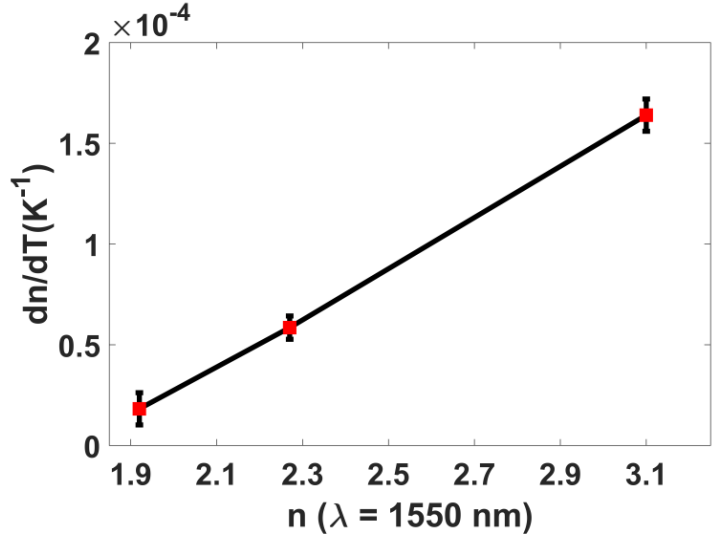

*Fig. 5. Enhancement of the thermo-optic coefficient and its relation to the refractive index with the corespoding uncertainties at $\lambda = 1550\ nm$*

**Table 2. Performance parameters of the MZI with 3 SiN$_x$ recipes for measuring thermo-optic coefficient**

| Sample # | w × h ($nm^2$) | $\Gamma_{SiN_x}$ | FSR (nm) | $n_g$ | $\frac{\Delta n}{\Delta T}(K^{-1})$ |
|---|---|---|---|---|---|
| S$_1$ | 800 × 550 | 59% | 11.5 | 1.84 | $(1.82 \pm 0.8) \times 10^{-5}$ |
| S$_2$ | 600 × 400 | 57% | 8.96 | 2.39 | $(5.85 \pm 0.5) \times 10^{-5}$ |
| S$_3$ | 400 × 320 | 69% | 5.94 | 3.96 | $(1.65 \pm 0.08) \times 10^{-4}$ |

## 5. SRN based Multi-Mode Interferometer Switch

To showcase the utility of the enhanced thermo-optic coefficient, high refractive index, and optical quality of SRN films such as S$_3$, we demonstrate a novel thermo-optically tuned MMI switch in SRN designed for operation in the C-band. A schematic of the SRN MMI switch is shown in Fig. 6, where the active waveguide cross section is similar to that of the MZI in the previous section. The MMI switch operates by the interference of two modes in a 320 nm tall and 600 nm wide multimode SRN waveguide. The multimode waveguide was designed to support fundamental and first order TE-like modes while retaining low waveguide propagation loss for both. Using Lumerical MODE simulations, the effective index and confinement factor are 2.42 and 84% for the fundamental and 1.62 and 31% for the first-order mode respectively. The multimode section of the device has 300nm wide inlet and outlet single mode waveguides with 300 nm width as shown in Fig. 6. The smaller width of the inlet and outlet waveguides insure single mode operation. The input waveguides (straight and bent) were designed so as to equally excite the fundamental and first order modes directly in the active waveguide section. Two S-bends merge symmetrically and meet in the center of the MMI waveguide to adiabatically excite the first order mode in the MMI section in a manner similar to that done for other MMI switches using silicon [22,23]. The time averaged intensity profile in the MMI as well as the excitation sections are shown in the inset of Fig. 6. The output portion of the switch, like the input section, consists of two single-mode, 45 μm bends which depart the MMI in an adiabatic fashion and form the two output ports. Depending on the effective optical length of the MMI for the two modes, the output of the MMI can be switched completely from one of the output ports to the other as shown in Fig. 6.

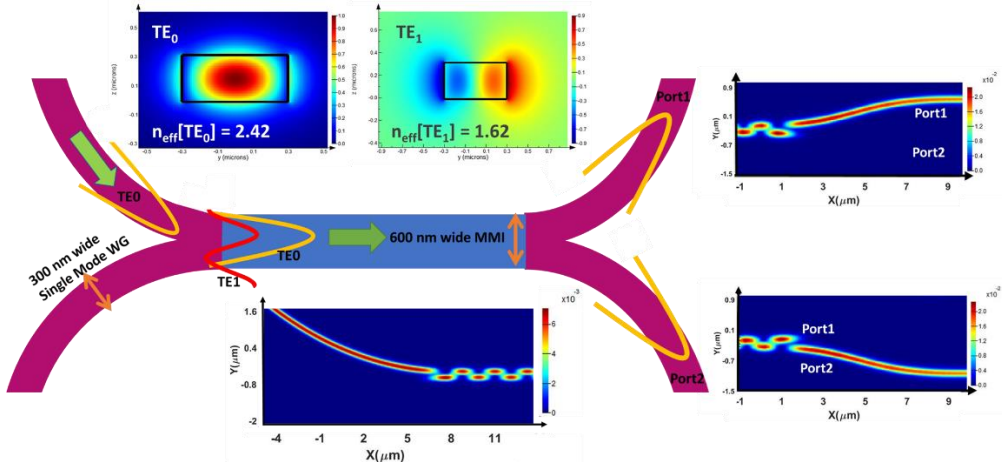

*Fig. 6. Schematic representation of the MMI device showing the single mode symmetrical bends at the input with the propagating fundamental mode into the 600 nm wide MMI section with the propagating multi-modes and the single mode output bend waveguides. Inset shown are the fundamental and the first order TE modes propagating in the MMI with their corresponding effective indices and the 2D FDTD simulation of the MMI excitation from the tapered waveguide section of 300 nm in addition to the energy transfer between the output ports.*

The characteristic beat length of the two MMI propagating modes is given by

$$L_B = \frac{\lambda}{2}\left[\frac{1}{n_{eff}(\text{TE0}) - n_{eff}(\text{TE1})}\right] \quad (3)$$

where $\lambda$ is the wavelength of the propagating optical field in vacuum, and $n_{eff}(\text{TE}_0)$ and $n_{eff}(\text{TE}_1)$ are the two modes' effective indices. A change in effective beat length of the two interfering modes can be actively induced by either changing the refractive index of the MMI waveguide material, or by changing the wavelength to achieve effective switching at the output ports. The former approach, in the case of SRN, can be induced by either utilizing its enhanced nonlinearity ($\chi^{(3)}$, using the well-known DC Kerr effect) [24] or exploit the enhanced thermo-optic effects as discussed below.

The fabrication of the SRN MMI switch, with an integrated Ni:Cr heater covering the multi-mode waveguide section, is carried out in a manner similar to that of the MZI in section 4. Optical switches with active SRN multimode waveguide sections as short as 800 μm was fabricated and characterized. The passive characteristics of the device, for an MMI section length of 800 $\mu m$, are shown in Fig. 7(a) as the measured normalized log-scale transmission at its two output ports as a function of wavelength. The characterization is carried out using the same setup as one used in section 4. As expected, the measured power at the two ports are conjugates of each other and show periodic peaks and valleys, with extinction ratios as high as 20 dB. The FSR of this response is measured to be approximately 3 nm. The FSR of such an MMI is expected to scale inversely with the length of the MMI and this is verified experimentally (not shown here) using fabricated devices of varying lengths. It should also be noted, that achieving such high extinction ratios validates the use of symmetric bends at the input for the efficient excitation of the two interfering modes. Subsequently thermo-optic switching was observed by measuring the transmission at both ports as a function of power consumed by the heater element. The measurements were carried out at a fixed wavelength, $\lambda$ = 1517.95 nm to maximize changes in output power as a function of index change. Fig. 7(b) shows the individually normalized powers at the two ports as a function of power consumed in the heater element. As evident from Fig. 7(b), upon application of power to the heater, the resulting change in temperature of the multi-mode waveguide, causes a change in output at both ports primarily as a result of thermo-optically induced change in refractive index of SRN. From these results one can observe that swings of up to 20 dB can be achieved in the output of individual ports and that the two output of the ports are exactly out of phase from each other, with the power required to completely switch the output from one port to another (defined as $P_\pi$) being ~50mW.

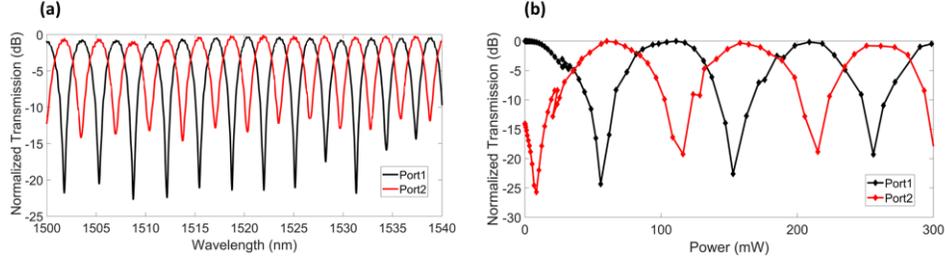

*Fig. 7. (a) Passive transmission spectra measured for 800 µm long MMI device for wavelengths of 1465 to 1570 nm (b) Transmission measured as a function of the power applied to the heater from 0 to 300 mW at $\lambda = 1517.95$ nm*

LUMERICAL DEVICE was used to perform transient thermal simulations to estimate the expected time-dependent switching characteristics of the device [25]. As expected, the thick oxide cladding (~2 $\mu m$) that separates the Ni:Cr heater from the waveguide and the 3 $\mu m$ bottom oxide separating the waveguide from the silicon substrate, cause the device to have relatively high rise and fall times ~50 $\mu s$. This implies a maximum switching rate of ~10KHz. It should be noted that this version of the SRN MMI switch was not optimized for switching speed. Dramatic reduction in rise and fall times can be obtained by switching to alternative heater configurations such as integrated doped silicon heaters, which allow for much lower separation between the waveguide and heating elements [25, 26]. While the current version of this MMI has a straight multi-mode waveguide 800 $\mu m$ in length, an optimized version can be designed with a combination of appropriately designed low-loss multimode straight and bend waveguides so as to allow for dramatic reduction in device footprint. In addition, the switching efficiency of this device can also be improved by a combination of detailed co-optimization of the active waveguide design and heater geometry and placement. The design and fabrication of such an optimized MMI thermo-optic switch will be reported in a future manuscript.

## 6. Summary and Conclusion

In this work we report experimental data on the thermo optic coefficient of PECVD deposited silicon nitride films, achieving a value as high as $(1.65\pm0.08) \times 10^{-4}$ K$^{-1}$ for a film with refractive index of 3.1. This is one of the highest reported thermo-optic coefficients in a PECVD deposited silicon nitride film to date and is comparable to that of crystalline silicon [27]. As expected, a clear correlation is experimentally demonstrated between the silicon concentration of the deposited silicon nitride films and their corresponding thermo-optic coefficients. We also demonstrate that these nitride films retain low optical loss, even at indices greater than 3. Further, we employ the highest thermo-optic coefficient to demonstrate an MMI based thermo-optic switch with a port-to-port switching power of 50 mW and extinction ratios as high as 20 dB. The combination of a high thermo-optic coefficient, low optical-loss, and ease of deposition using CMOS compatible tools such as PECVD, makes these SRN films ideal candidates for integrated thermo-optic applications.

## Funding

This project has been funded by the Office of Naval Research under Contract # N00014-18-1-2027

## Acknowledgment

We thank all UCSD's Nano3 cleanroom staff especially Dr. Maribel Montero, for their assistance in the preparation, fabrication, and characterization of the samples.

## Disclosures

The authors declare no conflicts of interest.